\documentclass[sigplan,screen,nonacm]{acmart}
\renewcommand\footnotetextcopyrightpermission[1]{} % removes footnote with conference information in first column
\usepackage{color}
\usepackage{listings}
\usepackage{cleveref}

\lstdefinestyle{mystyle}{
    basicstyle=\ttfamily\footnotesize,
    breakatwhitespace=false,
    breaklines=true,
    captionpos=b,
    keepspaces=true,
    numbers=left,
    numbersep=5pt,
    showspaces=false,
    showstringspaces=false,
    showtabs=false,
    tabsize=2
}

\AtBeginDocument{%
  \providecommand\BibTeX{{%
    \normalfont B\kern-0.5em{\scshape i\kern-0.25em b}\kern-0.8em\TeX}}}

\begin{document}

\title[Dr Wenowdis]{Dr Wenowdis: Specializing dynamic language C extensions using type information}

\author{Maxwell Bernstein}
\email{acm@bernsteinbear.com}
\orcid{0000-0003-3130-7059}
\affiliation{%
  \institution{Northeastern University}
  \streetaddress{}
  \city{Boston}
  \state{Massachusetts}
  \country{USA}
  \postcode{02115}
}

\author{CF Bolz-Tereick}
\email{cfbolz@gmx.de}
\orcid{0000-0003-4562-1356}
\affiliation{%
  \institution{Heinrich-Heine-Universität}
  \streetaddress{}
  \city{Düsseldorf}
  \country{Germany}}

\begin{abstract}
C-based interpreters such as CPython make extensive use of C ``extension''
code, which is opaque to static analysis tools and faster runtimes with JIT
compilers, such as PyPy. Not only are the extensions opaque, but the interface
between the dynamic language types and the C types can introduce impedance. We
hypothesise that frequent calls to C extension code introduce significant
overhead that is often unnecessary.

We validate this hypothesis by introducing a simple technique, ``typed
methods'', which allow selected C extension functions to have additional
metadata attached to them in a backward-compatible way. This additional
metadata makes it much easier for a JIT compiler (and as we show, even an
interpreter!) to significantly reduce the call and return overhead.

Although we have prototyped typed methods in PyPy, we suspect that the
same technique is applicable to a wider variety of language runtimes and
that the information can also be consumed by static analysis tooling.
\end{abstract}



%\keywords{Do, Not, Use, This, Code, Put, the, Correct, Terms, for,
%  Your, Paper}

\maketitle

\section{Introduction}

% glue:
% https://www.python.org/doc/essays/omg-darpa-mcc-position/
% https://numpy.org/doc/stable/user/c-info.python-as-glue.html
% https://dl.acm.org/doi/full/10.1145/3568973
% https://dl.acm.org/doi/10.1007/978-3-540-89965-5_24

One of the reasons for the success of dynamic languages such as Python and Ruby
is the ease with which they can interface to existing C libraries through the
use of C-implemented extension modules. Another common reason for writing
C extensions is to improve the performance when the dynamic language runtime
isn't fast enough in a hotspot. The downside of C extensions is that they
cannot easily be analyzed by static analysis tools together with the Python or
Ruby code that is calling into the library. The same problem plagues more
advanced dynamic language implementations with a JIT compiler because a call
into a C extension represents an optimization
barrier~\cite{chevalier_boisvert_yjit_2023}. For example, objects that may
otherwise be unboxed by the JIT~\cite{bolz_allocation_2011} now require boxing
for consumption by the C extension, only to often be immediately unboxed again
by the called C code.

As a motivating example, \Cref{lst:minimal-cext} shows a complete minimal
C extension module definition. The \verb|inc| function takes and returns a
Python \verb|int| object. It unboxes its argument, increments it, and re-boxes
the result. This type information is not available to the Python runtime; it's
implicit in the C argument processing wrapper code. Calling this function from
an optimizing Python implementation such as PyPy is thus very costly since it
requires a generic call path and the allocation of C data structures that
behave like the C extension expects them to.

In this paper we propose Dr Wenowdis~\cite{dr_wenowdis}, a very lightweight mechanism to expose some
amount of type and effect knowledge about the functions a C extension module
implements. We carry out this work specifically in the context of the CPython
C~API. We want to make it possible to incrementally add this knowledge to
existing libraries without having to do an invasive rewrite or introduce a new
dependency. We use the exposed information for improving
the performance of Python$\rightarrow$C calls using the PyPy JIT
compiler~\cite{bolz_tracing_2009}. The same type information can also
be used for type checking (e.g. in MyPy~\cite{mypy}) or static analysis.

For the example, we can add the type information that the function takes and returns a
C \verb|long| by writing the code in \Cref{lst:typedsigmacro}. This annotation is enough to
speed up calling the \verb|inc| function in PyPy by about 60 times, because it can
call the \verb|inc_impl| function directly, and optimize away the argument checking and
unboxing. The example will be discussed more thoroughly in \Cref{sec:capi} and 
\Cref{sec:typedsig}.

We present an early prototype that requires manual annotations with
very limited expressiveness but a more complete version could generate
the annotations automatically using a binding generator such as
Cython~\cite{behnel2010cython} or PyO3~\cite{pyo3}.

%CPython lives in PyObject land. PyPy lives in the JIT and in the JIT most
%objects are unboxed. When PyPy has to interact with C extensions it invents
%PyObjects. Some C extensions don't even really need PyObjects. Let's avoid
%making them.

\section{Background}

\subsection{CPython}

CPython is the reference implementation of Python. Older versions implement the Python
language by compiling it into a simple stack-based bytecode and running that in
a straightforward interpreter~\cite{barany2014python, power13falcon}. More
recent versions of CPython (from 3.11 onwards) use bytecode
quickening~\cite{brunthaler_ic, brunthaler_quickening,
brunthaler_ic_quickening, brunthaler_10ylater} to speed up bytecode
execution~\cite{lwn_faster_cpython}. The upcoming 3.13 release is probably
also going to contain a simple baseline JIT~\cite{cpython_jit} based on the
copy-and-patch approach~\cite{copy_and_patch}. CPython uses reference counting in
combination with a cycle-finding garbage collector to manage its memory.
CPython boxes all of its objects, including integers and floating point
numbers. It does not use pointer tagging or similar techniques.

\subsection{The CPython C~API}
\label{sec:capi}

While writing Python code is the normal way of interacting with the CPython
runtime, it is also possible to interact with it using its C~API. The C~API
is commonly used to create C extension modules, which expose new functions and
data types to Python code, that are implemented in C.\footnote{
The C~API also makes it possible to embed CPython into other projects, but this usecase is much less frequent and we won't discuss it in this paper.}
The C~API consists of a number of free functions %(\Cref{table:capi-functions})
and data types, %(\Cref{table:capi-data})
some of which are opaque to the
API client. It gives the tools to create Python objects, introspect them, call
Python functions, and more from C.

As an example, a minimal C extension can be found in \Cref{lst:minimal-cext}. 
In this C extension, \verb|PyInit_signature| sets up the module. It calls the C~API
function \verb|PyModule_Create|, which takes a description of the module it
wants to create: the \verb|PyModuleDef|. In the struct, we only define the
minimal features for this example: a name, documentation, and a method table.

\begin{figure}
\begin{lstlisting}[language=C, label=lst:minimal-cext, caption={A tiny C
extension, \texttt{signature}, exposing one function callable from Python,
\texttt{inc}. The function \texttt{PyInit\_signature} is called on first import.}]
#include <Python.h>

long inc_impl(long arg) { return arg+1; }

PyObject* inc(PyObject* module, PyObject* obj) {
  long l = PyLong_AsLong(obj);
  if (l == -1 && PyErr_Occurred()) return NULL;
  return PyLong_FromLong(inc_impl(l));
}

static PyMethodDef signature_methods[] = {
  {"inc", inc, METH_O, "Add one to a long."},
  {NULL, NULL, 0, NULL},
};

static struct PyModuleDef def = {
  PyModuleDef_HEAD_INIT, "signature", "doc", -1, signature_methods, NULL, NULL, NULL, NULL };

PyMODINIT_FUNC PyInit_signature(void) {
  return PyModule_Create(&def);
}
\end{lstlisting}
\end{figure}

\begin{figure}
\begin{lstlisting}[language=C, label=lst:typedsigmacro, caption={
Adding typing information to the minimal C extension in \Cref{lst:minimal-cext}.
}]
SIG(inc, LIST(T_C_LONG), T_C_LONG)
static PyMethodDef signature_methods[] = {
  TYPED_SIG(inc, inc, METH_O, "doc"),
  {NULL, NULL, 0, NULL},
};
\end{lstlisting}
\end{figure}

\verb|PyModule_Create| walks the method table (the array of
\verb|PyMethodDef|), creating \verb|PyCFunctionObject|s from the descriptions.
In this example, it creates a C function called \verb|inc| that takes one
argument (indicated by the flag \verb|METH_O|). When called, the C extension wrapper
code inside the Python runtime does argument count checking and then passes the
C function \verb|inc| the singular argument it needs.

Extension authors are typically required to write their own argument processing code.\footnote{
There is a CPython-internal preprocessor called Argument Clinic that automates
some of the work in writing argument processing code, but it is not meant for
external projects. We discuss it in \Cref{sec:clinic}.}
In this case,
we convert the argument from a  Python \verb|int| object to a C \verb|long| and raise 
an exception if that is not possible (this happens in \verb|PyLong_AsLong|).
Then we call the underlying C function, \verb|inc_impl|, and box up the result
for consumption in Python.

\subsection{PyPy}

PyPy is an alternative implementation of the Python language. PyPy is not implemented in C,
but in RPython, a statically typed subset of Python 2~\cite{ancona_rpython:_2007}. PyPy uses a
moving generational garbage collector for managing its memory. PyPy contains a tracing just-in-time
compiler to speed up the execution of Python code~\cite{bolz_tracing_2009}. To help the
JIT compiler generate better code, PyPy's object model is quite different than that of
CPython. In particular, Python instances are implemented using Self-style~\cite{chambers_efficient_1989} maps/hidden classes~\cite{bolz_impact_2015}.

\subsection{cpyext and its problems}

To allow PyPy to use the vast quantity of C extensions that exist for CPython,
PyPy has a compatibility layer for the CPython C~API, called
\textit{cpyext}~\cite{cuni_cpyext_2018}. It exposes (a subset of) the functions and
structs of the C~API.

Implementing this compatibility layer is quite challenging because CPython and PyPy function quite
differently. The CPython C~API exposes a number of internal implementation details of
CPython, most noticeably CPython's choice of reference counting for memory management.
%todo the example above does not demonstrate refcount semantics, maybe it should?
Handling Python objects from C requires the correct usage of \texttt{Py\_INCREF} and
\texttt{Py\_DECREF} everywhere in the C code.\footnote{This includes the
implicit runtime-owned wrapper code around C extension function calls.}
PyPy objects don't have a reference count field as the first word in each object,
and the PyPy GC would really like to be able to move objects as part of its minor collections. Therefore, PyPy creates tiny CPython-layout compatible structs for those 
of its objects that that are passed to C functions.

\begin{figure}[h]
% paste sequence diagram code into https://sequencediagram.org/
% note left of JIT: Python code
% JIT->>CPyExt: box integer
% CPyExt->>C Wrapper: Allocate PyObject*
% lifelinestyle C Wrapper #black:4:solid
% C Wrapper->>C Implementation: Unwrap PyObject*
% note right of C Implementation: Do work
% C Implementation->>C Wrapper: Allocate PyObject*
% C Wrapper->>CPyExt: Unwrap PyObject*
% CPyExt->>JIT: unbox integer
% note left of JIT: Python code
% note over C Wrapper:optimization barrier
\includegraphics[width=0.5\textwidth]{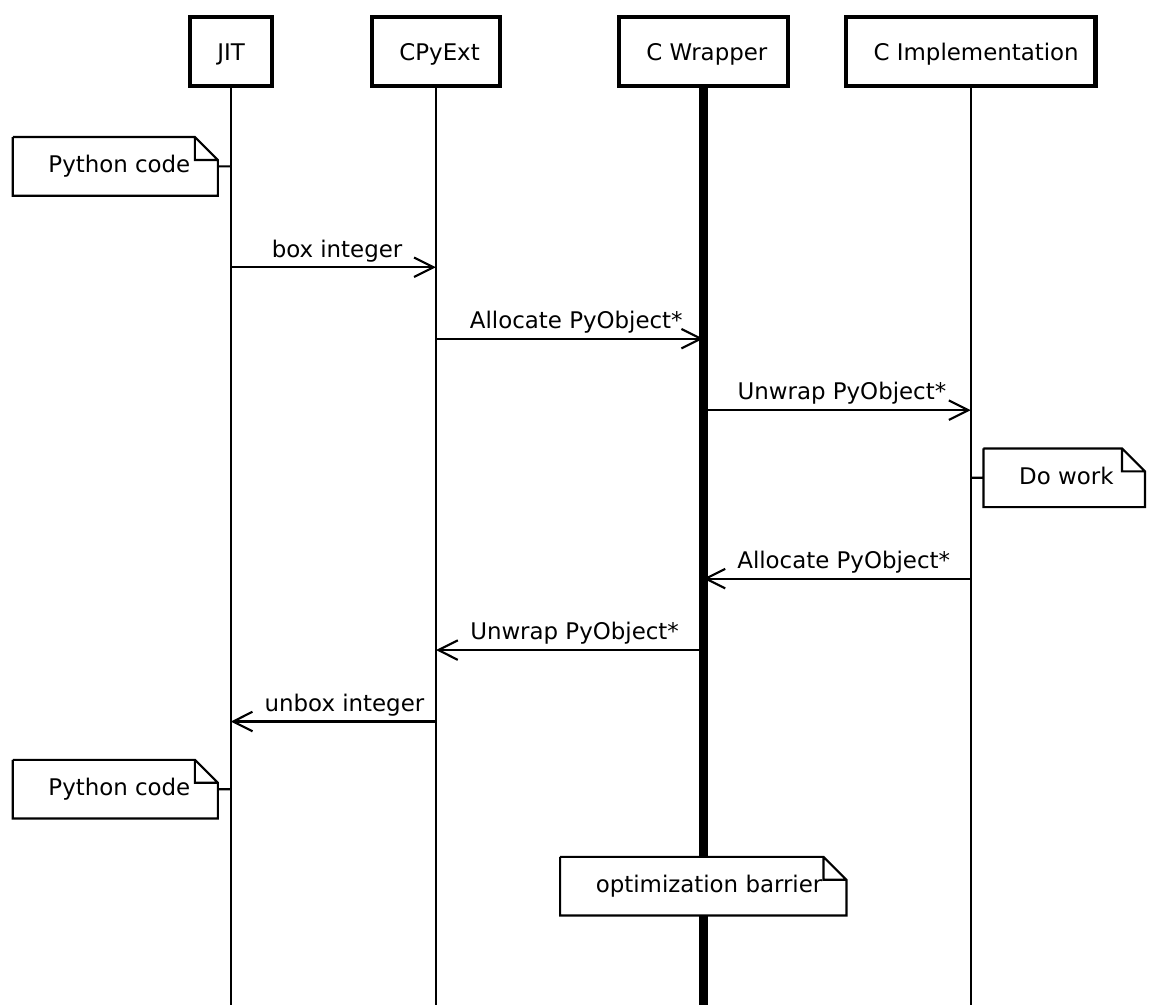}
\caption{Example call to from Python code optimized by the JIT to a C function, passing one integer argument. The diagram shows all the needed conversions in the process.}
\label{fig:sequence}
\end{figure}

Maintaining the link and converting between PyPy objects and CPython-layout compatible PyObjects is 
expensive. Every time PyPy calls a C function, we need to convert all of the function arguments to
PyObjects and then convert the result back to a PyPy object. This is particularly expensive for
boxed primitive types, because the C code will very likely just unbox them (with API functions such
\texttt{PyLong\_AsLong}) to work with the primitive values.
% TODO: Make this more prominent and potentially earlier
% laurie: the next paragraph is crucial
However, because all the argument
parsing and unboxing is done in C code and is therefore a black box, PyPy has no way to circumvent it.
This is the central problem that we want to address in this paper and is visualized in \Cref{fig:sequence}.

The problem becomes even more pronounced when PyPy's JIT is involved. The JIT will often compile
the Python code that calls a C-implemented function in a C extension module. The JIT 
infers the types of variables that are used in Python code at run-time. In the case of integers the JIT
will unbox them and store their integer values in machine registers~\cite{bolz_allocation_2011}.
In order to now pass such an unboxed integer as an argument to a C function, the JIT first has to recreate (i.e. allocate) a PyPy object,
which is then converted to a PyObject in order to pass it to the C code. The C code will then unbox
the value to work with the integer value itself. This kind of ping-pong between various 
representations is incredibly costly.

\section{Adding type information to C extensions}
\label{sec:typedsig}

We want to pass known type information from the C extension functions back
to Python. Doing this in a backwards compatible manner is non-trivial.
We describe some of the problems of doing so in this section.

\subsection{Exposing Type Information}

Our broad goal is to allow the runtime---the \emph{caller}---to make decisions
about argument type checking and unboxing instead of the C extension---the
callee. To do that, the runtime needs to know some type information and other
metadata about each C function.

In order to make this work, adding our type information must be
backwards-compatible both with PyPy and CPython. By this we mean that any
version of CPython or PyPy that does not understand the annotations should not
be tripped up by them: the C extensions should compile, load, and run just
fine. This is tricky because the Python C~API is not very extensible and also
because of some guarantees that Python makes about its C~API.

\subsection{The stable ABI}

The C~API exposed by CPython is consumed not only be extension authors, but
also by CPython developers.
Over the years, API clients have come to rely on details that have been exposed by the CPython C~API, 
such as struct sizes and structs fields that were originally intended for internal use only. Some source code has been 
completely lost to
time, and all people have are shared objects that call into C~API code. To support this
CPython has defined a stable binary interface (ABI).
% TODO(max): Figure out what wild .so IG uses that is completely lost to time.
% Perhaps also mention compilation matrices for different runtimes, versions,
% etc

Under the stable ABI contract, functions are not removed, functions do not add or remove
parameters,
and data types do not change size. There are a few additional minor restrictions.

If we're going to try and make existing C~API interactions faster under PyPy
with minimal effort, we need to find a way to add lightweight annotations to
methods. We can't change types, we can't change functions, and we can't make
people work too hard.

Because both the size of \verb|PyMethodDef|\footnote{
\url{https://docs.python.org/3/c-api/structures.html\#c.PyMethodDef}}
and the sizes and types of its
fields cannot change, we must smuggle in a pointer to more information stored
elsewhere.

\subsection{Sneaking in pointers}

We can at least signal that there \textit{is} additional information for a
\verb|PyMethodDef| by taking another bit in the \verb|ml_flags| bitset. We
propose the \verb|METH_TYPED| bit. When this bit is set, the PyPy extension
module loader knows to look for the extra type information.

\begin{lstlisting}[language=C]
struct PyMethodDef {
    const char  *ml_name;
    void        *ml_meth;
    int         ml_flags;
    const char  *ml_doc;
};
typedef struct PyMethodDef PyMethodDef;
\end{lstlisting}

Instead of the usual C string literal assigned to
\verb|ml_name|, we store the string in the \verb|PyPyTypedMethodMetadata|
struct and point \verb|ml_name| to that buffer. The name size we
chose is arbitrary. We then calculate an offset from that field to the beginning of the struct to use the added fields.

\begin{lstlisting}[language=C]
struct PyPyTypedMethodMetadata {
  int* arg_types;  // sentinel value of -1
  int ret_type;    // negative => can raise
  void* underlying_func;
  const char ml_name[100];
};
typedef struct PyPyTypedMethodMetadata PyPyTypedMethodMetadata;
\end{lstlisting}

A sample typed method looks like:

\begin{lstlisting}[language=C,escapechar=!]
int inc_arg_types[] = {T_C_LONG, -1};
struct PyPyTypedMethodMetadata inc_sig = {
  inc_arg_types, T_C_LONG, inc_impl, "inc",
};
static PyMethodDef signature_methods[] = {
  {inc_sig.ml_name, inc, METH_O!\colorbox{lightgray}{ | METH\_TYPED}!, "doc"},
  {NULL, NULL, 0, NULL},
};
\end{lstlisting}

To make this less irritating to write, we also provide macros to reach the form
that we already saw in \Cref{lst:typedsigmacro}. The macros also provide
another feature: backwards compatibility. Instead of doing \verb|#ifdef|
yourself for type signature feature detection, the macros do it for you. On
runtimes that support the \verb|METH_TYPED| flag, they emit signatures. On
runtimes that do not, they emit only standard C~API method metadata.

Once we know that the type information exists, we can use a trick from the
Linux kernel~\cite{kroah_hartman_2005} and read backwards from the
\verb|ml_name|:

\begin{lstlisting}[language=C]
PyPyTypedMethodMetadata*
GetTypedSignature(PyMethodDef* def)
{
  return (PyPyTypedMethodMetadata*)(def->ml_name - offsetof(PyPyTypedMethodMetadata, ml_name));
}
\end{lstlisting}

\section{Using type information in PyPy}

Once the argument and return type information is in place for a C extension,
this information can be used in cpyext. When we load a C extension module into
PyPy, we load the module's
methods. We check if each method has a \verb|METH_TYPED| flag set. If it does,
we find the metadata, build the signature, and store it on the internal method object.

When the function is called from PyPy, we first check
whether the called function has type information attached.
If that is the case, cpyext can use a fast path for implementing the call. The
arguments that are declared to be primitive types can be type-checked on the
PyPy side, without reboxing and subsequent conversion to \texttt{PyObject*}.
The call can then use the \texttt{underlying\_func} function pointer and
therefore skip the overhead of whatever Python calling convention the function
uses.

Being able to do the type checks for primitive arguments on the PyPy side (as
opposed to doing it in C) also meshes with PyPy's JIT type annotation, which means
the type check may not be required at all.

Last, instead of doing the slow and generic exception check, PyPy knows if the
function may never raise an exception---so it need not check---or what special sentinel
value to look for if it does raise. Functions can return \verb|NULL|, or
\verb|-1|, or something else depending on the return type to signal an error. This fast value
check removes the need for the full \verb|PyErr_Occurred()| call in the case
where the function did not signal that it raised. It is similar to CPython's
existing strategy for exception checking.

\section{Evaluation}

To evaluate our changes, we compare our modified PyPy against mainline PyPy,
CPython\footnote{We also tested the alpha release of the upcoming CPython 3.13
and it gave similar, if slightly slower results than CPython 3.12.}~\footnote{We
would have also liked to benchmark against the Cinder JIT, but the open-source
build for the JIT was broken at the time of writing.}, and GraalPy. We also
measure our modified PyPy \textit{with the JIT disabled} against mainline PyPy
with the JIT disabled.

% todo: We do not investigate the different kinds of C extension functions in the wild.

% \subsection{Measurement notes}
%
% % https://juliaci.github.io/BenchmarkTools.jl/stable/linuxtips/
% % https://easyperf.net/blog/2019/08/02/Perf-measurement-environment-on-Linux
%
% \verb|cpufreq-set -g performance| to turn off frequency scaling
%
% \verb|sudo cset shield -c 1 -k on| from \verb|cpuset| to prevent other
% processes from running on a core
% % But I am having trouble getting this working on my machine, maybe due to
% % systemd
%
% \verb|sudo cset shield -e julia -- benchmark.jl| to run on an isolated core
%
% % \verb|taskset --cpu-list 1 <cmd>| to run process on a core

% TODO(max): add a flag that will pass the module as a PyObject to the
% underlying C function

% TODO(max): Figure out if the extended error checking *before* C~API calls is
% necessary/correct or if it's just a waste of time and remove it; ask a
% CPython person (Lukasz?) or Matti

% TODO(max): specify that the typed function cannot be called with an exception
% set

At this early stage of our research we are only using some micro-benchmarks.
Every micro-benchmark is calling a C function many times in a hot loop.
The different benchmarks exercise different kinds of calls from Python into native code.
All the C function are themselves doing very little actual work. This means that
the performance is dominated by the overheads of the C~API and converting between the
different representations. The results therefore represent the performance ceiling: the
best possible improvements our approach can can make. They are not meant to represent
real-world code. The four microbenchmarks are:

\begin{itemize}
\item \verb|ffibench|, calling a \verb|METH_O| function with C types \texttt{long$\rightarrow$long}
\item \verb|objbench|, calling a \verb|METH_FASTCALL| function with C types \texttt{PyObject* $\rightarrow$ long $\rightarrow$ long}
\item \verb|idbench|, calling a \verb|METH_O| identity function with C types \texttt{PyObject* $\rightarrow$ PyObject*}.
\item \verb|idbench_exc|, also calling a \verb|METH_O| identity function with C types \texttt{PyObject* $\rightarrow$ PyObject*}, but this variant is annotated with the information that it can raise an exception.
\end{itemize}

\begin{figure*}[h]
    \centering
    \begin{minipage}{0.45\textwidth}
        \centering
        \includegraphics[width=1\textwidth]{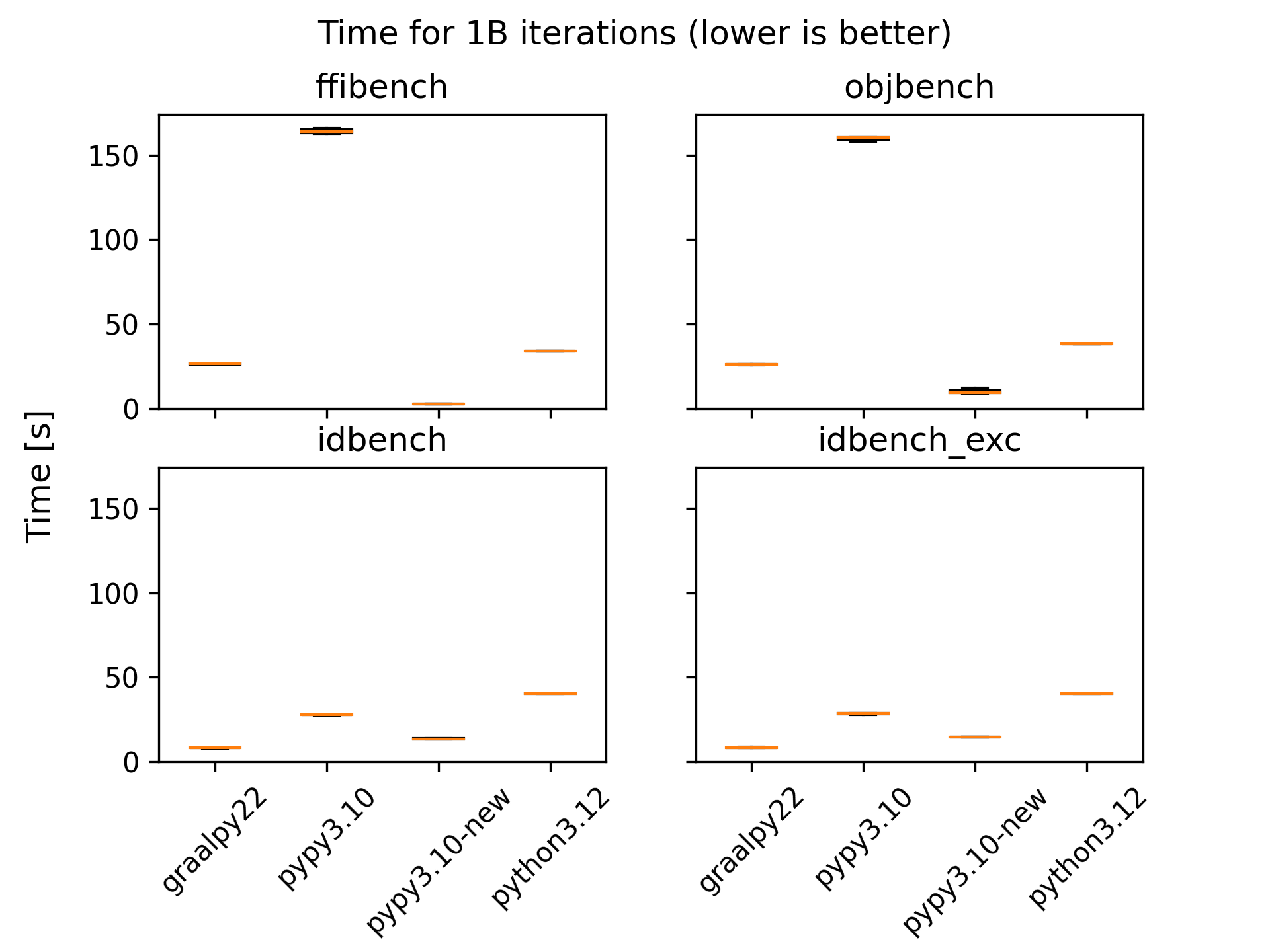} % first figure itself
        \caption{Benchmark results}
        \label{fig:bench}
    \end{minipage}\hfill
    \begin{minipage}{0.45\textwidth}
        \centering
        \includegraphics[width=1\textwidth]{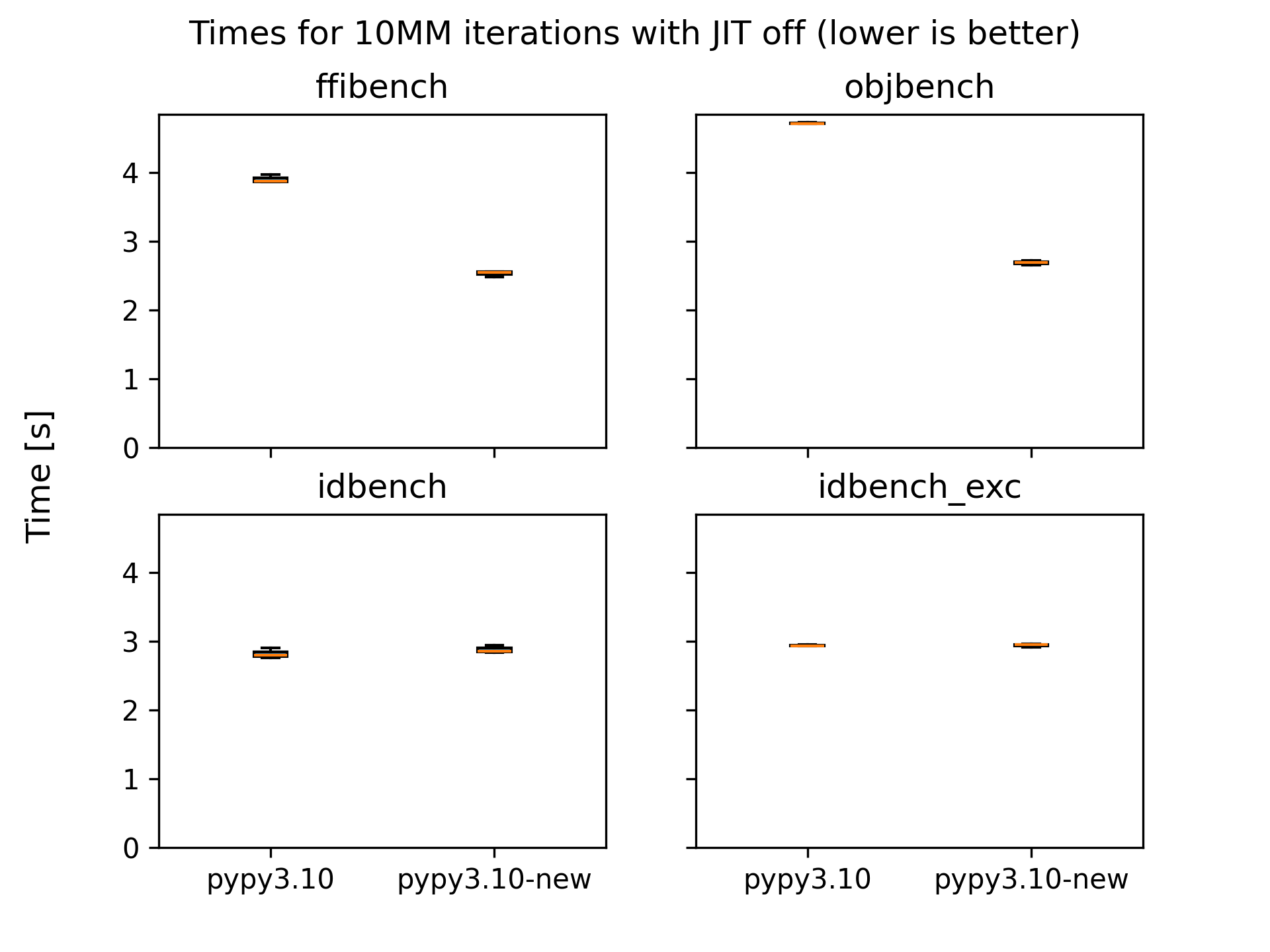} % second figure itself
        \caption{Benchmark for PyPy with JIT turned off}
        \label{fig:bench-multi}
    \end{minipage}
\end{figure*}

We run each benchmark 3 times for 1 billion iterations.\footnote{We specifically picked such high iteration counts to to give the Graal JIT enough time to warm up and have Sulong (\Cref{sec:sulong}) kick in for GraalPy 22. We omit GraalPy 23 because it took much longer 
than the other runtimes to finish and we killed the process. We hypothesize this is due to 
the removal of Sulong between versions 22 and 23. This makes GraalPy 23+ another good
candidate for using type information from C extensions.}
We then make a box and whisker plot of time taken for the three
runs. There is not much variance per runtime per benchmark. The results can be seen in \Cref{fig:bench}.

The best case benchmark for optimization is \verb|ffibench|. While the wrapper function deals in heap
allocated Python \verb|int|s, the underlying C function takes and returns a \verb|long|.
The type information lets
PyPy's JIT skip the overhead of creating \texttt{PyObject*} for the argument
and the return value completely. Our changes bring PyPy from slowest (about 164
seconds) to fastest (about 2.7 seconds).

In \verb|objbench| one of the C-functions arguments is unboxed and the 
other one requires allocating a \verb|PyObject| for each call. We do this to approximate a 
more realistic call. Not all calls to C extensions are going to be able to avoid all boxing.
In this benchmark, even though
PyPy is still allocating a \texttt{PyObject}, removing the overhead from
building an array of \texttt{PyObject} for the \verb|METH_FASTCALL| convention and from
boxing the result makes a big difference. Our changes bring PyPy 3.10 from
second slowest to fastest.

The third benchmark, \verb|idbench|, benchmarks the identity function. It is a
\verb|METH_O| function (one parameter) but this time it cannot type-specialize
its parameters or return value. This means that the runtime must box up the
argument into a \verb|PyObject| and unbox the return value the same way. Unlike
with \verb|METH_FASTCALL|, we are not eliminating any overhead for allocating
an argument tuple/array. Despite this, we cut PyPy's execution time in half and
are a close second place for time. GraalPy 22 is fastest, we suspect due to
Sulong being able to optimize the call to the identity function.

The last benchmark, \verb|idbench_exc|, is similar to
\verb|idbench| except that it can raise an exception. This means that the runtime must 
check first for an agreed-upon sentinel value (in this case, \verb|NULL|) and second check 
for an exception. Even though we are checking exceptions, we cut PyPy's
execution time in half and are a close second place for time.

Last, we ran all four benchmarks in PyPy with the JIT turned off
(\Cref{fig:bench-multi}). Our changes still improve PyPy performance because the runtime need not allocate the arguments array of \verb|PyObjects| for each
call to a \verb|METH_FASTCALL| function; it knows how many arguments to
pass and can pass them in registers when calling the underlying function.
Additionally, the runtime can skip slow thread-local storage (TLS) lookups in exception checking for C extension functions that
cannot raise Python exceptions.

\vspace{-\topsep}

\section{Prior work}

\citet{monat_multilanguage_sa} built a multi-language static analysis platform
called Mopsa. They analyze several open-source libraries and their unit tests
and find that multiple Python$\leftrightarrow$C calls happen per unit
test (ranging from 2.7 to 51.7). Their analysis could
potentially be repurposed to generate type information for C extensions (they
note that they would like to infer standard library type information using
their techniques). They note that the analysis is limited by use of
hard-to-analyze Python libraries and imprecision in the C analysis. This work
is expanded in \citet{monat_thesis}.

Similarly, \citet{hu2021pyctype} describe PyCType, which automatically infers the argument types for functions exposed by Python C extension in order to find bugs. This information should be usable for runtime performance improvements.

\citet{tsai_improving_2013} use the LLVM JIT to speed up the performance of JNA callbacks in the Java Hotspot Server VM \cite{tsai_improving_2013}. Their approach yields 8-16\% performance improvements and does not apply to \emph{calling} C functions from Java (only the other way around).

\citet{li_tan_jet, li_tan_turbojet} find bugs in Java Native Interface (JNI)
modules related to exception-checking. Their tools, JET and TurboJet, implement
a static analysis to find missing declarations for checked exceptions.
Automatically finding and annotating C extensions that do not raise exceptions
could help improve run-time performance with less manual work.

The HPy project~\cite{hpy} is a complete re-design of the CPython C~API from the ground up. 
One of its main goals is to move away from reference counting being visible in the API. 
However, it still does nothing to solve the problems discussed in this paper.

% \citet{felgentreff_towards_2015}

%I guess it's not really normal FFI but instead this weird other thing where you load C %extension modules. This means you don't in general have the lowest-level signature %information

%I don't know when people use FFI (CFFI, for example) compared to a C extension

%\subsection{Cinder and Pyjion}

%Python comes with an extensive suite of builtin modules. These modules are
%loaded mostly the same as any other C extension module is loaded. JIT compilers
%built on top of CPython such as Cinder and Pyjion want to be able to optimize
%calls to these builtin functions. Because there is no type information in
%upstream CPython, they each have to maintain internal tables of information
%about the builtins.

%While the tables are useful for one kind of optimization (better flow typing
%leading to more specialized code generation), they falls short in two important
%ways:

%First, the tables are maintained manually. While the names and types are
%unlikely to change version-to-version, they are not guaranteed to be stable. In
%addition, CPython upstream adds new functions in many releases and these must
%be manually tracked by the Cinder and Pyjion maintainers.

%Second, the tables do not provide a function pointer to the underlying C
%function. Calls to these functions must still go through the normal CPython
%argument passing and returning machinery, which should not be necessary in a
%JIT.

%PyPy does not have tables like the ones found in Cinder and Pyjion because its
%builtins are written entirely in RPython, a language that is interpretable by
%the JIT.

\vspace{-\topsep}
\subsection{Sulong}
\label{sec:sulong}

Sulong~\cite{rigger_bringing_2016} is an self-optimizing interpreter based on the Truffle framework~\cite{wurthinger_one_2013} for LLVM~\cite{lattner_llvm:_2004} bitcode. It can be used to speed up calling from Python into C extensions by JIT-compiling both the Python and the C code in the same compilation unit using the Graal JIT compiler. This gets rid of most of the conversion overhead, but has the downside that called C function is also running on top of Sulong, which is slower than using a well-tuned static C compiler for the core algorithms of extensions.

\subsection{Argument Clinic}
\label{sec:clinic}

The CPython runtime has an internal tool called the Argument
Clinic~\cite{hastings_clinic} that takes as input descriptions of C functions
and generates argument processing code and documentation strings for them. The
clinic is used only to generate standard library code in CPython.

\subsection{Static Python}

The Cinder project\footnote{\url{https://github.com/facebookincubator/cinder/}} is
perhaps most widely known for its JIT compiler, but it also includes a compiler
and runtime for a statically-typed dialect of Python known as Static Python
(SP)~\cite{lu_gradual_2022}. The SP compiler contains primitives for declaring,
resolving, and calling C functions directly from Static Python code. The SP
compiler produces typed bytecode, which allows the JIT to compile specialized,
zero-overhead calls to these C functions. See \Cref{lst:sp-abs} for an example.

\begin{lstlisting}[language=Python, label={lst:sp-abs}, caption={Taken from the
  \href{https://github.com/facebookincubator/cinder/blob/2d2c1d68e47a553aae6dd5786cc3493723046136/cinderx/RuntimeTests/hir_tests/hir_builder_native_calls_test.txt}{Cinder
  test suite}. In this example, the
\texttt{abs} function is declared as a stub to be loaded from \texttt{libc.so}.
It takes an unboxed (C) \texttt{int32} and returns the same. The \texttt{test}
function takes a boxed (Python) \texttt{int}, unboxes, calls \texttt{abs}, and
re-boxes.}]
from __static__ import native, int32, box

@native("libc.so.6")
def abs(i: int32) -> int32:
    pass

def abs_wrapper(i: int) -> int:
    j: int32 = int32(i)
    return box(abs(j))
\end{lstlisting}

The JIT uses \texttt{dlsym} and the typed SP bytecode to build a
\texttt{NativeTarget}: a function pointer, return type, and argument types.

Annotating declaration of existing C functions is a manual process. Also, the SP
compiler does not allow passing \texttt{int}s into the unboxed \texttt{abs}
function; callers must explicitly unbox.

\vspace{-\topsep}
\vspace{-\topsep}

\section{Conclusion}

We have shown that adding type information to C extensions can make them faster
under the PyPy JIT. We have also shown that the techniques improve performance on PyPy even with the JIT compiler turned \emph{off}.

Type information specialization is effective
even in an interpreted context and potentially even without
unboxed objects. We believe that this technique is not limited to PyPy and
can be adopted by other dynamic language runtimes.
For example, in runtimes such as Skybison and MicroPython with efficient
representations for small objects (such as tagged integers in pointers), the
runtime need not allocate a \verb|PyObject| for each argument and return value;
it can use the efficient representation directly \cite{skybison,micropython}.

% other proposals: Add \_Py\_AttachSig(string\_name, sig) symbol and call it in
% module/type/... init. runtimes can support it how they like.
%
% or: instead of declaratively specifying args and return, could generate a small
% bytecode for an arg processing virtual machine. maybe.

\vspace{-\topsep}

\section{Future work}

Given how promising our early results are, we would like to build out support for more complex signatures, such as support
for C strings and primitive types wider than 64 bits. We would also like to
support a more expressive declaration language, such as the one used in the
Argument Clinic.

% A production-ready implementation of this work would also contain version
% information so that type annotation data layout changes do not prevent
% optimization, or worse, cause silent data corruption.

In the future, we would like to see these type annotations automatically
emitted by binding generators. Projects such as Cython and PyO3 already have
machinery for generating wrapper code for C functions, and therefore have
sufficient knowledge about the C function types.

Such binding generators also have more insight into the effects that happen
inside a native function. For example, Cython may be able to statically
guarantee that a function does not raise an exception, need to acquire the GIL,
or something else. Reducing the set of effects from ``all effects possible''
could aid the PyPy optimizer.

%In a more distant future, we would like to see more advanced ahead-of-time
%compilers such as Cython and Mypyc generate code that is interpretable by PyPy.
%This would make the otherwise opaque C code visible to the JIT compiler and
%even remove the need for the type signatures in the best case.

\begin{acks}
To Sarah, for proposing the title. To Kate McKinnon, for ongoing comedic genius.
\end{acks}

\bibliographystyle{ACM-Reference-Format}
\bibliography{sample-base}

\end{document}